\documentclass[preprint,preprintnumbers,amsmath,amssymb,aps]{revtex4}

\usepackage{graphicx}
\usepackage{dcolumn}
\usepackage{bm}
\usepackage{enumerate}
\usepackage{epsfig}

\begin{document}
\title{A Simple General Solution for Maximal Horizontal Range of Projectile Motion}
\author{Boris Bu\v{s}i\'{c}}
\affiliation{%
 Ksaver 173, Zagreb, Croatia
}%

\email{borisb@hi.hinet.hr}

\date{\today}

\begin{abstract}
 A convenient change of variables in the problem of maximizing the horizontal
range of the projectile motion, with an arbitrary initial vertical
position of the projectile, provides a simple, straightforward solution.
\end{abstract}

\maketitle

\section{Introduction}

\begin{figure}[t]
\centering
\includegraphics[angle=0,
          width=0.65\linewidth]{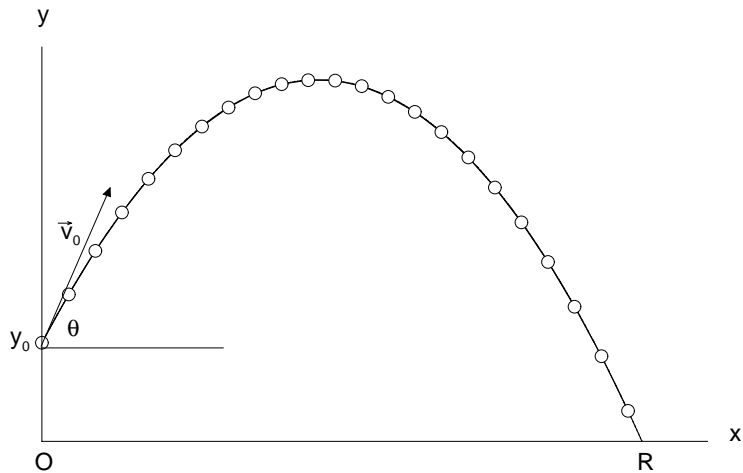}
\caption{The trajectory  of a body projected from the height $y_0$ with an
  initial velocity $\vec{v_0}$ at an angle $\theta$ with respect to the horizontal.
  $R$ is the horizontal range of the motion.  }
\label{BD}
\end{figure}

A clear, concise formulation of the general solution
for maximal horizontal range of the projectile motion
is considered non-trivial.
Consequently much effort has been spent to formulate a solution that would
be understandable to as wide audience as possible. The problem may be stated
as follows: 
\begin{enumerate}[]
\item
Find the maximal horizontal range, $R$, 
of a projectile launched  with an initial speed $v_0$
from the point with the vertical coordinate $y_0$.
(Figure \ref{BD}) 
\end{enumerate}

Generally the known solutions are divided into those that do 
and those that do not make use of calculus.
Somewhat misleadingly the former are commonly qualified as simple.
The categorization according to whether a solution starts from the vector
or scalar general solution of projectile motion seems more appropriate. 

In the former one starts from
\begin{equation}
\begin{split}
  \vec{r} &= \vec{r_0} + \vec{v_0}t + \frac{1}{2} \vec{g} t^2 \\
  \vec{v} &= \vec{v_0} + \vec{g} t
\end{split}
\label{vs}
\end{equation}
and employs ``ingenious application of vector algebra, involving both
dot and cross products" (citation from Thomsen \cite{thom}). 
An example of a solution from this category is given by Palfy-Muhoray and
Balzarini \cite{palf}.
While it does not involve calculus the solution can hardly be considered simple. 
The use of vector algebra as the means for the solution seems to be 
be inspired by the paper of Winans \cite{win} who 
provided a solution by using quaternion multiplication.

In the latter approach one starts from the more familiar parametric representation
of the projectile trajectory
\begin{equation}
\begin{split}
x  &= v_{0} t \cos{\theta}  \\
y  &= y_0 + v_{0} t \sin{\theta} - \frac{1}{2} gt^2.
\end{split}
\label{base}
\end{equation}
The calculus-based solutions (e.g. \cite{porter,licht}) proceed with
the substitutions,
\begin{equation}
  y = 0, \qquad x = R, \qquad t = \frac{R}{v_0 \cos{\theta}},
\label{sub}
\end{equation}
and look for the maximum of the resulting function $R=R(\theta)$ by setting
\begin{equation}
  dR/d\theta = 0.
\label{calc}
\end{equation}
 Lichtenberg and Wills \cite{licht}, to whom this method is usually attributed,  
maximize
\begin{equation}
   R = \frac{v_0^2}{g} \cos{\theta} \bigg[\sin{\theta} + 
       \sqrt{\sin^2{\theta} + \frac{2gy_0}{v_0^2}} \bigg].
\label{lw}
\end{equation}
While the method is straightforward and therefore may be considered simple,
the function (\ref{lw}) is quite complicated and the application of the 
necessary condition for the extremum of a function of one variable (\ref{calc}) 
leads to a very involving calculation. This may be the main reason
to look for the solutions that do not involve calculus.

The substitutions (\ref{sub}) accompanied with the use of the trigonometric identity
\begin{equation}
  \frac{1}{\cos^2{\theta}} = 1 + \tan^2{\theta}
\end{equation}
give
\begin{equation}
  0 = y_0 +  R \tan{\theta} - \frac{g R^2}{2 v_0^2} (1 + \tan^2{\theta}),
\label{dep}
\end{equation}
which is the basis for a couple of interesting non-calculus-based solutions.
Thomsen \cite{thom}, for example, considers 
Equation (\ref{dep})
as a quadratic equation in $\tan{\theta}$. The requirement that the solutions of
the equation be real then leads to a solution of the problem. 
Ba\'{c}e \textit{et al.}\cite{bace} take the same view of Equation (\ref{dep})
but obtain the solution through
a more pointed observation that the maximal horizontal range, $R_{max}$, is 
realized for the unique initial direction of the projectile.

Some additional solutions are discussed in Brown \cite{brown}, who also 
gives his own, non-calculus-based solution based
on the solution of a related problem of the range of a projectile 
launched down an incline.

\section{Change of Variable Solution}

Instead of focusing on some explicit relation between the relevant variables,
e.g. $R = R(\theta)$ or $R = R(t)$, one can
notice that the Equation (\ref{dep}) can be cast in the form where 
$R^2$ is expressed as a quadratic function of $R \medspace \tan{\theta}$,
\begin{equation}
R^2 = - (R \medspace \tan{\theta})^2 + 2 \frac{v_0^2}{g} (R \medspace \tan{\theta})
       + 2 \frac{v_0^2 y_0}{g}.
\label{my}
\end{equation}
This observation makes both, calculus-based and purely algebraic, solutions
simple.

In the former case 
\begin{equation}
  dR^2/d(R\tan{\theta}) = 0 
\end{equation}
is equivalent to
\begin{equation}
     R_{max} \tan{\theta_{max}} = \frac{v_0^2}{g}.
\label{tmax}
\end{equation}
Equation (\ref{my}) then gives 
\begin{equation}
 R_{max} = \frac{v_0}{g}  \sqrt{v_0^2 + 2 g y_0},
\label{rsol}
\end{equation}
which in turn, inserted into Equation (\ref{tmax}), leads to 
\begin{equation}
  \tan{\theta_{max}} = \frac{v_0}{\sqrt{v_0^2 + 2 g y_0}}.
\label{tansol}
\end{equation}
This completes the solution.

If, on the other hand, an algebraic solution is preferred one can transform 
Equation (\ref{my}) into
\begin{equation}
  R^2 - \bigg [ 2 \frac{v_0^2 y_0}{g} + \frac{v_0^4}{g^2} \bigg ] =
   - \bigg[ R \medspace \tan{\theta} - \frac{v_0^2}{g} \bigg ]^2
\label{sol}
\end{equation}
and notice that the right-hand side is non-positive. Thus
\begin{equation}
   R^2 - \bigg [ 2 \frac{v_0^2 y_0}{g} + \frac{v_0^4}{g^2} \bigg ]
          \leq  0,
\end{equation}
and
\begin{equation}
 R_{max}^2 = \frac{v_0^2}{g^2} \big ( v_0^2 + 2 g y_0 \big )
\end{equation}
at
\begin{equation}
  R_{max} \medspace \tan{\theta_{max}} = \frac{v_0^2}{g}.
\end{equation}
The last two expressions then lead to (\ref{rsol}) and (\ref{tansol}).

\section{Conclusion}

The present solution of the problem of maximizing the horizontal range
of the projectile motion is based on the widely applicable technique of
change of variables. 
Although this may be implicit in the algebraic variant
of the solution,
both variants may serve
as an illustration of the usefulness of the technique in simplifying
otherwise complicated calculations.

\acknowledgments{
Many thanks to M.Ba\'{c}e and Z.Naran\v{c}i\'{c} for bringing the problem to my
attention and urging me to publish this solution.
}

\bibliographystyle{apsrev}
\bibliography{bib}

\input{bibnames.sty}
\begin{thebibliography}{7}
\expandafter\ifx\csname natexlab\endcsname\relax\def\natexlab#1{#1}\fi
\expandafter\ifx\csname bibnamefont\endcsname\relax
  \def\bibnamefont#1{#1}\fi
\expandafter\ifx\csname bibfnamefont\endcsname\relax
  \def\bibfnamefont#1{#1}\fi
\expandafter\ifx\csname citenamefont\endcsname\relax
  \def\citenamefont#1{#1}\fi
\expandafter\ifx\csname url\endcsname\relax
  \def\url#1{\texttt{#1}}\fi
\expandafter\ifx\csname urlprefix\endcsname\relax\def\urlprefix{URL }\fi
\providecommand{\bibinfo}[2]{#2}
\providecommand{\eprint}[2][]{\url{#2}}

\bibitem[{\citenamefont{J.S.Thomsen}(1984)}]{thom}
\bibinfo{author}{\bibnamefont{J.S.Thomsen}}, \bibinfo{journal}{Am.J.Phys.}
  \textbf{\bibinfo{volume}{52}}, \bibinfo{pages}{881} (\bibinfo{year}{1984}).

\bibitem[{\citenamefont{P.Palfy-Muhoray and D.Balzarini}(1982)}]{palf}
\bibinfo{author}{\bibnamefont{P.Palfy-Muhoray}} \bibnamefont{and}
  \bibinfo{author}{\bibnamefont{D.Balzarini}}, \bibinfo{journal}{Am.J.Phys.}
  \textbf{\bibinfo{volume}{50}}, \bibinfo{pages}{181} (\bibinfo{year}{1982}).

\bibitem[{\citenamefont{J.G.Winans}(1961)}]{win}
\bibinfo{author}{\bibnamefont{J.G.Winans}}, \bibinfo{journal}{Am.J.Phys.}
  \textbf{\bibinfo{volume}{29}}, \bibinfo{pages}{623} (\bibinfo{year}{1961}).

\bibitem[{\citenamefont{D.B.Lichtenberg and J.G.Wills}(1978)}]{licht}
\bibinfo{author}{\bibnamefont{D.B.Lichtenberg}} \bibnamefont{and}
  \bibinfo{author}{\bibnamefont{J.G.Wills}}, \bibinfo{journal}{Am.J.Phys.}
  \textbf{\bibinfo{volume}{46}}, \bibinfo{pages}{546} (\bibinfo{year}{1978}).

\bibitem[{\citenamefont{W.S.Porter}(1977)}]{porter}
\bibinfo{author}{\bibnamefont{W.S.Porter}}, \bibinfo{journal}{Phys.Teach.}
  \textbf{\bibinfo{volume}{15}}, \bibinfo{pages}{358} (\bibinfo{year}{1977}).

\bibitem[{\citenamefont{M.Ba\'{c}e et~al.}(2002)\citenamefont{M.Ba\'{c}e,
  S.Iliji\'{c}, and Z.Naran\v{c}i\'{c}}}]{bace}
\bibinfo{author}{\bibnamefont{M.Ba\'{c}e}},
  \bibinfo{author}{\bibnamefont{S.Iliji\'{c}}}, \bibnamefont{and}
  \bibinfo{author}{\bibnamefont{Z.Naran\v{c}i\'{c}}},
  \bibinfo{journal}{Eur.J.Phys.} \textbf{\bibinfo{volume}{23}},
  \bibinfo{pages}{409} (\bibinfo{year}{2002}).

\bibitem[{\citenamefont{R.A.Brown}(1992)}]{brown}
\bibinfo{author}{\bibnamefont{R.A.Brown}}, \bibinfo{journal}{Phys. Teach.}
  \textbf{\bibinfo{volume}{30}}, \bibinfo{pages}{344} (\bibinfo{year}{1992}).

\end{thebibliography}
\end{document}